\documentstyle[preprint,aps,epsf,rotate]{revtex}
\def\ffrac#1#2{\textstyle{#1\over#2}\displaystyle}
\begin{document}
\draft
\preprint{Version 21 Feb 2003} 
\title{Lecture on Branched Polymers and Dimensional Reduction}
\author{John Cardy}
\address{Institute for Advanced Study, Princeton NJ 08540 \\
 and University of Oxford, Department of Physics -- Theoretical
         Physics, 1 Keble Road, Oxford OX1 3NP, U.K.\footnote{Address for
correspondence} \\
         and All Souls College, Oxford.}
%
%\date{\today}
%
\maketitle
\begin{abstract}
This is a pedagogical account of the some of the recent results of Brydges
and Imbrie, described from the point of view
of Grassmann integration. Some simple extensions are pointed out.
\end{abstract}
\subsection*{Historical Introduction}

The subject of dimensional reduction, in the context considered in this
lecture, has a long and rather peculiar history.
The strangest aspect of all is that the simplest formulation
of the results has been found only very recently. 

That one physical theory in $d$ space dimensions should be related to another
in $D=d-2$ dimensions was first suggested in the context of the field-theoretic
formulation of the critical 
behaviour of the Ising model in a quenched random magnetic field\cite{IM,Young}. 
It was noticed
that the most singular Feynman diagrams all had the form of trees before 
averaging over the quenched randomness. When this 
was done, it turned out that the diagrams were the same as those
for the critical Ising model without any random field, in
two fewer dimensions. Thus, no further work was needed -- one should be able
to take,
for example, the known results for the ordinary Ising model in $D$
dimensions, and find the critical behaviour of the random
field Ising model in $d=D+2$ dimensions. In particular, since the lower
critical dimension (at and below which there is no phase transition) of
the ordinary Ising model is $D=1$, this should imply that the 3-dimensional
random field Ising model has no phase transition. 

Unfortunately this beautiful idea is wrong. In particular it contradicts
a very simple argument of Imry and Ma\cite{IM} that the lower critical dimension
should be two. However, it took several years of theoretical and experimental
confusion and contradictory papers before Imbrie\cite{Imbrie} proved that Imry and Ma
were right. 

During this time Parisi and Sourlas\cite{PS1} came up with a startling explanation of
how and why dimensional reduction works: supersymmetry. At that time 
supersymmetry was thought to be the preserve of particle physics and string 
theory, so their argument was not widely understood. In their paper was also the
germ of the reason why simple dimensional reduction might not work for the
random field Ising model: the tree diagrams were seen to be the perturbative
solution to the classical field equations in the presence of the random
external field. If the solution of these equations was unique and given
by the sum of the diagrams, then all should have been well. But it was
easy to see that, at low enough temperatures, the solution was not unique,
and that the one 
with the lowest energy was not the perturbative one. Work is still
continuing on how, possibly, to rectify this situation, but this seems a very
difficult problem.

Meanwhile, a few years later, Parisi and Sourlas\cite{PS2} came up with another
ingenious application of their ideas. The problem known under various
versions as branched polymers, lattice trees, or lattice animals
(to be decribed later) had been attracting some attention.
Lubensky and Isaacson\cite{LH} had formulated a rather complicated field theory for
this problem, had come to the conclusion that the upper critical dimension
(above which mean field theory is valid) is $d=8$, and had noticed that
the first terms in the $\epsilon$-expansion of the critical exponents below
this were the same as those of the so-called Yang-Lee edge singularity
in $D=d-2$ dimensions. This is the problem of an Ising model in a purely
imaginary magnetic field, and is described by a simple scalar field theory
with a cubic interaction and a purely imaginary coupling. Parisi and
Sourlas reformulated the model of Lubensky and Isaacson in a simpler way,
and then
used their supersymmetry to explain why the dimensional reduction happened.
This time, everything was all right. The numerically measured exponents of
lattice animals and lattice trees in $d=3$ dimensions are given by the
exactly known Yang-Lee exponents in $D=1$. A non-perturbative proof of
dimensional reduction in a supersymmetric field theory was given\cite{JC}.
It was understood that the main reason for the failure of dimensional
reduction in the random field Ising model does not apply here, essentially
because the tree diagrams \em are \em the branched polymer 
configurations\cite{Shapir}.
But none of these arguments were completely watertight, largely because
it seemed necessary to ignore so-called irrelevant terms which spoiled
the supersymmetry, but were believed not to affect the critical behaviour.

In fact, the Yang-Lee critical theory is nowadays strongly believed\cite{LF} to be
in the same universality class as an even
simpler problem: a classical gas with short-range repulsive
interactions in the grand canonical ensemble. As a function of the
fugacity $z>0$, the grand partition function is sum of positive terms
and has no singularity nor any zeroes. But in the complex $z$ plane the
closest singularity to the origin lies on the negative axis at $z=-z_c$, and
the critical behaviour near this is believed to be independent of
the particular details of the interaction (so it is known as the `universal
repulsive gas singularity'), and it is also
believed to be the same as that of the Yang-Lee problem at the critical
imaginary magnetic field. Thus the branched polymer problem in $d$
dimensions was believed to be related to the universal repulsive gas
singularity in $D=d-2$ dimensions. But once again these arguments relied
on the neglect of irrelevant terms in the appropriate field theories.

It took Brydges and Imbrie\cite{BI1} to realise that most of this
theoretical undergrowth could be cut away - that field theory is not needed
at all! There is a simple and physical model for branched polymers in
$d$ dimensions for which dimensional reduction to a repulsive gas
is rigorous and exact,
not just in the critical region (which in this context means very large
polymers) but for all values of the fugacity. 
The arguments do use supersymmetry, in
a very beautiful way, and give compelling evidence of how important it
is in theoretical physics to learn how to walk (do problems with
finitely many particles) before one tries to run (field theory and
beyond.)

The aim of these lectures is to present these ideas in as simple a manner
as possible, using the ideas of Grassmann integration nowadays familiar
to many theoretical physicists. I shall, however, quickly recall all the
ingredients necessary for the argument.

\subsection*{Classical gas and cluster expansion.}

Consider a classical gas of point particles, in some large $D$-dimensional
domain $\Omega$,
interacting through a rotationally symmetric two-body potential $V$.
The grand partition function is
\begin{equation}
\Xi(z)=\sum_{N=0}^\infty Z_Nz^N
\end{equation}
where
\begin{equation}
Z_N={1\over N!}\int_\Omega\prod_{j=1}^Nd^Dr_j e^{-\sum_{j<k}V(r^2_{jk})}
\end{equation}
where $r^2_{jk}\equiv(r_j-r_k)^2$ and we work in units where $kT=1$.
As the volume $|\Omega|$ of the domain goes to infinity, we expect,
if $V$ is sufficiently short-ranged, that $\Xi(z)\sim e^{p(z)|\Omega|}$,
where $p(z)$ is the pressure. 

For our purposes it is easier to consider the mean density
$n=z(d/dz)p(z)$, which has the expansion
\begin{equation}
\label{n}
n(z)= (\Xi(z))^{-1}\sum_{N=1}^\infty {z^N\over (N-1)!}
\int_\Omega\prod_{j=1}^Nd^Dr_j \delta^{D}(r_1)e^{-\sum_{j<k}V(r^2_{jk})}
\end{equation}
The cluster expansion follows by writing $e^{-V(r^2)}=1-G(r^2)$, so that
$e^{-\sum_{j<k}V(r^2_{jk})}=\prod_{j<k}\big(1-G(r^2_{jk})\big)$, and
expanding everything in powers of $G$. Note that if $V$ is short-ranged,
so is $G$. For the density, all the contributions which diverge as
$|\Omega|\to\infty$ cancel between the numerator and denominator,
and we are left with a sum of connected cluster diagrams. The coefficient
of $z^N$ is a sum of connected graphs $\cal G$, 
each of which has $N$ nodes labelled
by $(r_1,\ldots,r_N)$, of which $r_1$ is fixed to be at the origin. 
The value of the graph is given by integrating over all the other $r_j$
the product of the $-G(r^2_{jk})$ over all connected edges $(jk)$ of the
graph. An example is shown in Fig.~1. 

Note that the integral corresponding to a particular graph $\cal G$
can be performed by inserting the integral representation
\begin{equation}
\label{rep}
G(r^2_{jk})=\int_0^\infty f(\alpha_{jk})e^{-\alpha_{jk}r_{jk}^2}d\alpha_{jk}
\end{equation}
on each edge: the integral over the $r_j$ then has the form
\begin{equation}
\int\prod_jd^Dr_j\delta^{(D)}(r_1)
e^{-\frac12\sum_{jk}\sum_{\mu=1}^Dr^\mu_i{A_{\cal G}}_{jk}r^\mu_k}
\end{equation}
where $A_{\cal G}$ is a matrix which depends on the $\{\alpha_{jk}\}$, and
on $\cal G$, but not on $D$. 
The only feature of this we shall need is that it factorises into
$D$ integrals over each component of the vectors $r_j$, and
by the rules of gaussian integration it
has the form 
\begin{equation}
\label{form}
(2\pi)^{(N-1)D/2}(\det A_{\cal G})^{-D/2} 
\end{equation}

\subsection*{Grassmann integration}
Grassmann coordinates behave in many ways like ordinary coordinates,
except that they anticommute. For our purposes, we are going to
consider a space with just two Grassmann coordinates, called $\theta$
and $\overline\theta$. It is useful to consider these as analogues
of complex commuting coordinates $z=x+iy$ and $\overline z=x-iy$, so that the
squared distance between two points $(\theta_j,\overline\theta_j)$
and  $(\theta_k,\overline\theta_k)$ is the analogue of 
$(\overline z_j-\overline z_k)(z_j-z_k)$, that is 
$|\theta_{jk}|^2\equiv(\overline\theta_j-\overline\theta_k)
(\theta_j-\theta_k)$. 

We need the notion of integrating some function $f(\overline\theta,
\theta)$ over all space. Note that because ${\overline\theta}^2=
\theta^2=0$, the Taylor expansion of any such function terminates:
$f=A+\overline B\theta+C\overline\theta+D\overline\theta\theta$.
One important property of integration over all space is that we may
shift the integration variables by constants, so that
\begin{equation}
\int d\overline\theta d\theta f(\overline\theta+\overline\kappa,\theta+\kappa)
=\int d\overline\theta d\theta f(\overline\theta,\theta)
\end{equation}
This can only be satisfied if 
\begin{equation}
\label{GI}
\int d\overline\theta d\theta\,1=
\int d\overline\theta d\theta\,\overline\theta=
\int d\overline\theta d\theta\,\theta=0
\end{equation}
but we are allowed to choose the normalisation of
\begin{equation}
\int d\overline\theta d\theta\,\overline\theta\theta=\Lambda
\end{equation}

We also have to learn about gaussian Grassmann integration: the main rule
is that if $B$ is an $N\times N$ matrix
\begin{equation}
\int\prod_jd\overline\theta_jd\theta_j e^{-\frac12\sum_{jk}
\overline\theta_jB_{jk}\theta_k}=(-\ffrac12\Lambda)^N\det B
\end{equation}
We shall not try to prove this in general\cite{IZ}. Let us note the simplest
case $N=1$: 
\begin{equation}
\int d\overline\theta d\theta e^{-\frac12 B\overline\theta\theta}
=\int d\overline\theta d\theta(1-\ffrac12 B\overline\theta\theta)
=-\ffrac12\Lambda B
\end{equation}
\subsection*{Supersymmetric classical gas}
We are now going to consider a classical gas which lives in superspace.
Points in superspace are labelled by $d$ commuting coordinates, the components
of $r$, and two anticommuting coordinates $(\overline\theta,\theta)$ as above.
The squared distance in superspace is 
\begin{equation}
R^2_{jk}=r^2_{jk}+|\theta_{jk}|^2=
(r_j-r_k)^2+(\overline\theta_j-\overline\theta_k)(\theta_j-\theta_k)
\end{equation}
and our classical gas with $N$ particles has grand partition function 
$\Xi_{\rm SS}(z)=\sum_{N=0}^\infty Z_{N,{\rm SS}}z^N$, where
\begin{equation}
\label{SS}
Z_{N,{\rm SS}}= {1\over N!}\int\prod_{j=1}^Nd^dr_jd\overline\theta_jd\theta_j
e^{-\sum_{j<k}V(R^2_{jk})}
\end{equation}
Here `SS' stands for superspace, or for `supersymmetric'. The supersymmetry
in this case is under super-rotations 
in the $d+2$ dimensional space. Note that there
are many different kinds of supersymmetry. This one is rather different
from that considered in most particle physics applications, and is also
different from the global supersymmetries used in quenched random systems
in condensed matter physics.

Actually, the partition function for this supersymmetric gas is not
very interesting: since $V$ depends only on the relative coordinates, the
integration over the center-of-mass anticommuting coordinates gives zero,
by virtue of (\ref{GI}), except when $N=0$. Thus $\Xi_{\rm SS}(z)=1$.
This is typical of supersymmetric theories. However, the mean density
is non-trivial:
\begin{equation}
\label{nSS}
n_{\rm SS}(z)=\sum_{N=1}^\infty{z^N\over(N-1)!}
\int\prod_{j=1}^Nd^dr_jd\overline\theta_jd\theta_j
\delta^{(d)}(r_1)\delta(\overline\theta_1)\delta(\theta_1)
e^{-\sum_{j<k}V(R^2_{jk})}
\end{equation}
where $\delta(\overline\theta_1)\delta(\theta_1)$ just sets
$\overline\theta_1=\theta_1=0$ in whatever expression follows it.

We are going to evaluate (\ref{nSS}) in two different ways: first we shall
show that it is identical to the same expression (\ref{n}) for the
ordinary classical gas in $D=d-2$ dimensions, then we shall evaluate it
in another way to show that it gives branched polymers in $d$ dimensions.

\subsection*{Dimensional reduction}
As before, we can make a cluster expansion of (\ref{nSS}) by writing
$e^{-V}=1-G$. In this case, only connected graphs survive because any
disconnected pieces vanish on integrating over their center-of-mass 
anticommuting coordinates. Using the integral representation (\ref{rep}),
a given graph $\cal G$ now contributes (before doing the $\{\alpha_{jk}\}$
integrations)
\begin{eqnarray*}
&&\int\prod_jd^dr_jd\overline\theta_jd\theta_j
\delta^{(d)}(r_1)\delta(\overline\theta_1)\delta(\theta_1)
e^{-\frac12\sum_{jk}\big(r_jA_{jk}r_k+\overline\theta_jA_{jk}\theta_k\big)}
\\&&\qquad\qquad=(2\pi)^{(N-1)d/2}\,(\det A_{\cal G})^{-d/2}\,
(-\ffrac12\Lambda)^{N-1} (\det A_{\cal G})
\end{eqnarray*}
and we see that this is identical to (\ref{form}), with $D=d-2$,
as long as we take
\begin{equation}
\Lambda=-1/\pi\quad.
\end{equation}
To summarise, we have
\begin{equation}
n_{\rm SS}^{(d,2)}(z)=n^{(d-2)}(z)
\end{equation}
In a sense, the integrations over the
anticommuting coordinates cancel those over two of the commuting ones.
This again is a result typical of supersymmetry.

\subsection*{Branched polymers}
We haven't yet tried to define the term `branched polymer.' Physically,
it is an object consisting of nodes connected by linear segments which are
chains of monomers. Each node has a specific degree, the number of
segments meeting at that node, which can either be $1$ or any integer
$\geq3$. By extending this defintion, we can think of the connections between
each pair of monomers on the linear sections as nodes of degree $2$.
Each branched polymer has the topology of a tree, with 
a fixed number of monomers
on each segment or branch. 
It is embedded in $d$-dimensional euclidean space, and
there are assumed to be interactions between the monomers which (a) represent
the fact that neighbouring monomers on the tree
are confined within a typical radius $a$,
and (b) there is a steric repulsion between all other pairs of monomers
which inhibits them from approaching each other closer than a distance also
$O(a)$. The branched polymer is assumed to be in thermal equilibrium with
some kind of solvent, so that is explores its allowed phase space, each 
configuration occurring with a probability $\propto e^{-{\rm energy}/kT}$,
according to the laws of statistical mechanics.

For enumeration purposes it is often simpler to allow the nodes to lie
at the sites of a regular lattice, of spacing $O(a)$, and then to count,
with equal weights, all allowed configurations of a given total number
of monomers $N$. Depending on whether loops are allowed or not, this gives
the problems of lattice animals or lattice trees. Numerical work suggests
that, for large $N$, these models exhibit a kind of critical behaviour,
characterised by universal critical exponents. For example, the 
typical size $R\sim N^\nu$, and the number of distinct animals (or trees)
which are not related by lattice translations $p_N$, goes like 
$N^{\alpha-1}\mu^N$, where $\mu$ is model-dependent, but the exponents
$\nu$ and $\alpha$ are universal. The connection with conventional critical
phenomena is made more clear by considering the generating function
$g(z)=\sum_Np_Nz^N$, where $z=z_c=\mu^{-1}$ is the critical point, near
which $g(z)\sim (z_c-z)^{1-\alpha}$.

We shall show that the supersymmetric model (\ref{SS}) gives rise
to a continuum model of branched polymers which in fact is much more realistic
than these lattice models.

\subsection*{From the supersymmetric gas to branched polymers}
Go back to the formula (\ref{nSS}) for the density and write, for each pair
of particles 
\begin{equation}
e^{-V(R^2_{jk})}\equiv P(R^2_{jk})=P(r^2_{jk}+\overline\theta_{jk}\theta_{jk})
=P(r^2_{jk})+Q(r^2_{jk})\overline\theta_{jk}\theta_{jk}
\end{equation}
where $Q(r^2)=P'(r^2)$. 
So we can write the integrand in (\ref{nSS}) as a sum of $2^{N(N-1)/2}$
terms, in each of which for each pair $(jk)$ we choose either the term
$P(r^2_{jk})$ or the term $Q(r^2_{jk})\overline\theta_{jk}\theta_{jk}$.
Each term can be represented by a subgraph of the complete graph connecting
all $N$ nodes, in which the edge $(jk)$ is included if we choose the
second term, and excluded if we choose the first. See Fig.~\ref{fig2}.
Now consider performing the integrations over the anticommuting coordinates.
For the same reason as above, any subgraphs not connected to the origin
will vanish.
Moreover, it is fairly easy to see that any subgraphs containing cycles 
will also vanish, since this will always involve some anticommuting
coordinate being raised to a power $\geq 2$. Thus the surviving subgraphs
all have the form of \em rooted connected trees\em, rooted because the particle
at the origin is privileged. It also easy to see that all possible such
rooted trees with $N$ nodes occur (we allow nodes of degree 2.) For a given
tree ${\cal T}$ we may then relabel the coordinates, starting from the origin.
The integration over the anticommuting coordinates on
one particular edge of the subgraph is then of the form
\begin{equation}
\int d\overline\theta d\theta Q(r^2)\overline\theta\theta=(-1/\pi)Q(r^2)
\end{equation}
We therefore have
\begin{equation}
n^{(d,2)}_{\rm SS}(z)
=\sum_{N=1}^\infty z(-z/\pi)^{N-1}N{\cal Z}_N
\end{equation}
where
\begin{equation}
\label{ZT}
{\cal Z}_N=
\sum_{{\cal T}}{1\over N!}\int\prod_{j=1}^Nd^dr_j\delta^{(d)}(r_1)
\prod_{j\sim k}Q(r^2_{jk})\prod_{j\not\sim k}P(r^2_{jk})
\end{equation}
where $j\sim k$ means that $j$ and $k$ are neighbouring nodes on ${\cal T}$,
and $j\not\sim k$ means that they are not, and each pair $(jk)$ is counted
just once.

For suitable potentials $V$, we can interpret this as the partition
function for a branched polymer with the topology of $\cal T$: 
$P(r_{jk}^2)=e^{-V(r^2_{jk})}$ is the Boltzmann weight representing the
steric repulsion between different parts of the polymer, and $Q(r^2_{jk})$
then is the weight for two neighbouring nodes, which keeps them a distance
$O(a)$ apart. If the original gas had only repulsive forces, then $P'\geq0$
so that $Q$ may be interpreted as a weight.
Fortunately it is quite easy to arrange a suitable function $Q$. An example
is illustrated in Fig.~\ref{fig3}. 
A special case is to choose $V(r^2)=V_0>0$ for
$r<a$ and zero for $r>a$. Then 
\begin{equation}
Q(r^2)=(1-e^{-V_0})\delta(r^2-a^2)=(2a)^{-1}(1-e^{-V_0})\delta(r-a)
\end{equation}
If we take the limit $V_0\to\infty$ corresponding to a hard core
repulsive gas, then the corresponding branched polymer model may be thought
of as consisting of balls of radius $a/2$, with neighbouring balls on the
tree constrained to touch each other, and a hard core repulsion between
the rest.

Putting together all the pieces of the puzzle, we have shown that the
generating function for rooted branched polymers in $d$ dimensions 
is related to the mean density of a repulsive gas in $D=d-2$ dimensions
by
\begin{equation}
z(d/dz)p(z)=n(z)=
\sum_{N=1}^\infty z(-z/\pi)^{N-1} 
N{\cal Z}_N
\end{equation}

By integrating with respect to $z$, we finally end up with a relation
between the pressure $p(z)=\lim_{|\Omega|\to\infty}|\Omega|^{-1}\ln\Xi(z)$
of the classical gas and the generating function for branched polymers:
\begin{equation}
\label{main}
p(z)=-\pi \Sigma(-z/\pi)
\end{equation}
where
\begin{equation}
\Sigma(\zeta)=\sum_{N=1}^\infty {\cal Z}_N\zeta^N
\end{equation}
Eq.~(\ref{main}) is the main result of Brydges and Imbrie\cite{BI1}.

\subsection*{Simple examples}
\subsubsection{$d=2$}
When $d=2$ then $D=0$, a particularly simple case, since the universe
is all at one point\cite{Calvino}. In that case, we do not have to worry about
the thermodynamic limit either. The grand partition function is 
\begin{equation}
\label{Xi}
\Xi(z)=\sum_{N=0}^\infty{z^N\over N!}e^{-\frac12N(N-1)V(0)}
\end{equation}
since if there are $N$ particles at the origin, there are 
$\frac12N(N-1)$ interactions between them. The remarkable thing is that
the branched polymer partition function in $d=2$ depends
only on the value of the potential $V(r)$ at the origin. This is a 
consequence of the supersymmetry. If we relax the condition that $Q=P'$,
it is no longer true. Although the sum in (\ref{Xi}) cannot be performed
explicitly, it can be shown\cite{CR} that all the zeroes of
$\Xi(z)$ lie on the negative real axis and are distinct. The closest
one to the origin determines the large $N$ behaviour of ${\cal Z}_N$,
and gives rise to the branched polymer singularity. 

In the hard core case $V(0)\to\infty$, there can only be zero or one
particles, so we have
\begin{equation}
p(z)=\ln\Xi(z)=\ln(1+z)
\end{equation}
so that
\begin{equation}
\Sigma(\zeta)=-(1/\pi)\ln(1-\pi\zeta)
\end{equation}
and\footnote{Note that this differs by factors of $2$ from equation (1.4) of
Ref.~\cite{BI1}, since we have $Q=(1/2a)\delta(r-a)$, while these authors
take $Q=\delta(r-1)$.}
\begin{equation}
{\cal Z}_N=\pi^{N-1}/N
\end{equation}
It is very hard to see how this simple result could be obtained by other means.
While it is easy to check for $N=1,2$, it already is non-trivial for $N=3$,
while for $N\geq4$ the integrals involved in each contribution appear intractable.
\subsubsection{$d=3$}
Interacting classical gases in $D=1$ can generally be solved exactly only in the
case of infinite hard core repulsion, i.e.~$V(0)\to\infty$. We can think of
the particles as rods which we now take to be of unit length.
If the total length is $L$ and we
use periodic boundary conditions then since configuration space has
volume $(L-N)^N$, we have $Z_N=(L-N)^N/N!$ 
A classic calculation in statistical mechanics then shows that, in the
thermodynamic limit, $p=n/(1-n)$
(the denominator is just the excluded volume term in van der Waals' equation!) 
and so $z(dp/dz)=n=p/(1+p)$. Integrating, we find an implicit equation for
the pressure: $z=pe^p$. Remarkably, this can be solved explicitly to find
$p(z)$ as a power series in $z$ [Hint: write $\oint p(z)dz/z^{N+1}$
as a similar contour integral over $p$]
\begin{equation}
p(z)=\sum_{N=1}^\infty {(-N)^{N-1}\over N!}z^N
\end{equation}
Using the main result (\ref{main}) this expansion once again gives some
highly nontrivial identities for sums of certain integrals in three
dimensions. 

If we are just interested in the
branched polymer singularity note that $z(p)$ has a quadratic turning point
at $z=-z_c=-e^{-1}$, so that $p(z)$ has a square root branch point there.
This is consistent with the square root singularity at the Yang-Lee edge in
$D=1$: it is an easy exercise in transfer matrices
to solve the nearest neighbour Ising model in a purely imaginary field,
and check this.

\subsection*{Correlation functions}
Although we have so far considered only the mean density, it is easy
to generalise dimensional reduction to correlation functions:
\begin{equation}
\langle n(r_1)n(r_2)\ldots n(r_p)\rangle_{\rm SS}^{(d,2)}
=\langle n(r_1)n(r_2)\ldots n(r_p)\rangle^{(d-2,0)}
\end{equation}
which is valid \em as long as \em the points $r_j$ all lie in the
$(d-2)$-dimensional subspace, and at $\overline\theta=\theta=0$.
If we now expand the left hand side in
terms of $P$ and $Q$ as before we find the generating function for
$p$ \em different \em branched polymers, which exclude each other's volume,
rooted at the points $r_j$.
Of course, for $p=2$ the correlation function enjoys full rotational
invariance in superspace. Thus if we take $r_1=0$ and $r_2=r$ and we let
\begin{equation}
G(r^2+\overline\theta\theta)
=\langle n(0)n(r,\overline\theta,\theta)\rangle_{\rm SS}^{(d,2)}
\end{equation}
then $G(r^2)$ is the same as the density-density correlation function
for the $D$-dimensional ordinary gas. Writing now
$G(r^2+\overline\theta\theta)=G(r^2)+\overline\theta\theta G'(r^2)$,
the first term gives as before the correlation function between two
different but interacting branched polymers, while, after performing
the Grassmann integrations, we see that the second term, proportional
to $G'(r^2)$, gives the density-density correlation function on 
a \em single \em branched polymer. That these are related is a consequence
of the supersymmetry. This is all spelled out in full detail in Ref.~\cite{BI2}.
Since for $D=1$ the correlation function can in some cases be computed exactly,
so can that for branched polymers in $d=3$. This was first carried out using 
the Yang-Lee model by Miller\cite{Miller}.

\subsection*{Extensions}
I'll finish by briefly mentioning some simple extensions of this
beautiful idea.
\subsubsection*{Bells and whistles} 
It is possible to decorate the nodes of the branched polymer with other
degrees of freedom, e.g. to consider that they are of different species,
and that the interactions between them can depend on this. Under dimensional 
reduction this becomes a classical gas with the same decoration. It would be nice
if we could find a different universality class at negative fugacity. Field
theory suggests that this might be the case (we choose a theory with a 
$\phi^{2n+1}$ interaction with $n\geq2$, rather than $\phi^3$) but investigations 
so far\cite{CR}
have shown that, at least for physical values of the interactions, the universal
repulsive singularity is robust, and nothing new happens.
\subsubsection*{Adsorption at a wall or a plane} 
The astute listener should have realised that all that is needed for dimensional
reduction to work is rotational invariance in two of the $d$ commuting dimensions,
so that integrals over these can be cancelled by the two anticommuting ones.
Thus, for $d\geq3$, we are free to add an arbitrary potential $U(x)$ depending on
one of the commuting coordinates $x$. The generating function for branched polymers 
in $d$ dimensions in the presence of this potential will be related to that
of a $D$-dimensional classical gas in the same manner as (\ref{main}).

As an example take a hard core repulsive gas in
$d=3$, and $e^{-U(x)}=1+\lambda\delta(x)$, with $\lambda>0$,
representing an attractive potential on the plane $x=0$. A simple calculation
then shows that the density at a point close to, but not on, the plane
is modified to
\begin{equation}
n(z)={n_0(z)\over 1+\lambda n_0(z)}
\end{equation}
where $n_0(z)$ is the unperturbed density.
For small $\lambda$, the nearest singularity to the origin is still at
$z=-z_c$, the $d=3$ branched polymer singularity. However, for large
enough $\lambda$ it instead comes from the vanishing of the denominator,
and is generally a simple pole. This is the branched polymer singularity
in $d=2$. In this regime the branched polymer is confined to the vicinity
of the plane $x=0$.
The critical value of $\lambda$ at which these singularities collide
represents an adsorption transition. The case of an attractive wall, where
$U=+\infty$ for $x<0$, may be dealt with similarly.
\subsubsection*{Dimensional reduction by 4}
We do not have to stop at adding one pair of anticommuting coordinates -
for example we can add another, in which case we get dimensional reduction
by four. But the equivalent $d$-dimensional model then involves interaction
weights proportional to $P$, $P'$ and $P''$, and the last typically
has to change sign, so is unphysical.
\subsection*{Intriguing Connections} If one takes the scaling limit of
the expression (\ref{Xi}) as $z\to -z_c$ and $V(0)\to 0$, one finds 
something proportional to an Airy function. The Airy integral arises in many
contexts, but another was recently found by Richard, Guttmann and
Jensen\cite{RGJ}, who argued that it occurs in the scaling limit of self-avoiding
closed loops in the plane, weighted by their length and the area enclosed within.
The parameters conjugate to these are the fugacity, analogous to $z$, and the
(negative) pressure inside the loop, 
which plays the same role as $V(0)$ in the branched
polymer generating function. Using the field theory arguments of Parisi and
Sourlas\cite{PS2}, I have argued elsewhere\cite{JCSAP} why an Airy function 
is expected in this context. This also explains why the fractal dimension of
self-avoiding loops is $\frac43$. But it would be very interesting to provide a
direct connection between branched polymers and self-avoiding loops -- is perhaps
the external perimeter (suitably defined) 
of a large branched polymer in the same universality class
as that of ordinary self-avoiding loops?

\subsection*{Acknowledgements}
Over the years I have benefited from conversations on this subject with N.~Sourlas,
Y.~Shapir and J.~Miller, and more recently with J.~Imbrie and R.~Rajesh.
These lectures were written while the author was a member of the Institute for
Advanced Study.
He thanks the School of Mathematics and the School of Natural Sciences
for their hospitality.
This stay was supported by the Bell Fund, the James D.~Wolfensohn Fund,
and a grant in aid from the Funds for Natural Sciences.

Since these notes were written, another review by Imbrie\cite{JI} has appeared.

\begin{figure}
\label{fig1}
\centerline{
\epsfxsize=5cm
\epsfbox{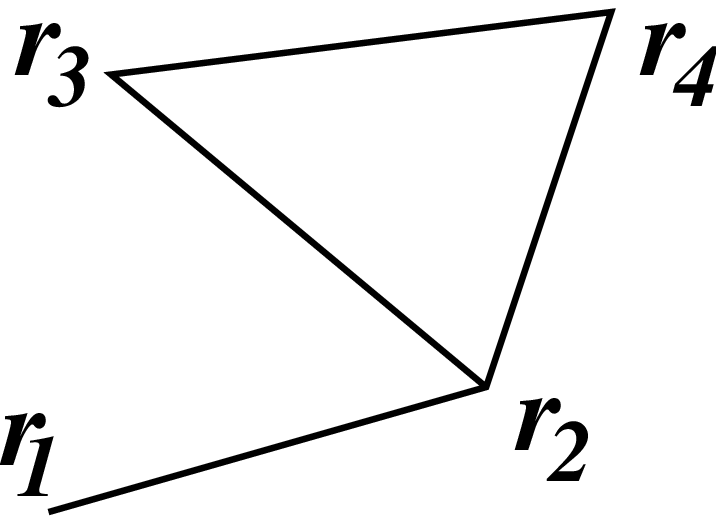}}
\caption{A typical cluster diagram for the mean density.}
\end{figure}
\begin{figure}
\label{fig2}
\centerline{
\epsfxsize=8cm
\epsfbox{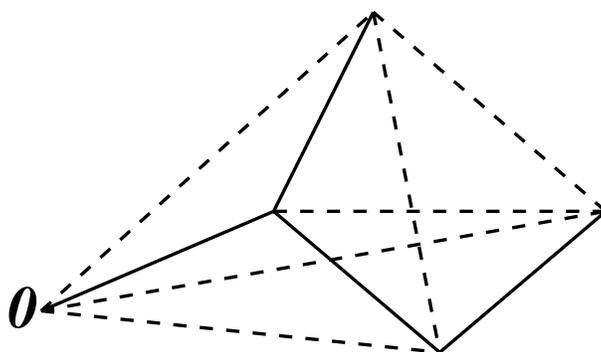}}
\caption{A branched polymer configuration. Solid lines correspond to the
nerest neighbour interaction $Q$, and dashed lines to repulsive interactions
$P$.}
\end{figure}
\begin{figure}
\label{fig3}
\centerline{
\epsfxsize=8cm
\epsfbox{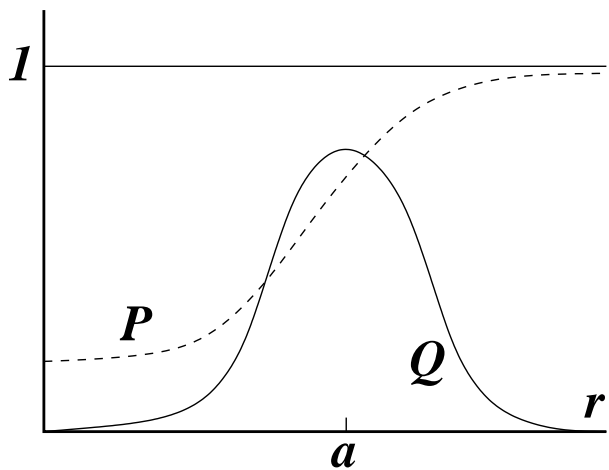}}
\caption{Typical forms for $P$ (dashed curve) and $Q=P'$ (solid).}
\end{figure}
\end{document}